\documentclass[preprint,superscriptaddress,showpacs,nobalancelastpage]{revtex4}
\usepackage{graphicx,amsmath,amssymb}

\begin{document}

\title{Dirac-screening stabilized surface-state transport in a topological insulator}

\author{Christoph Br\"une}
\affiliation{Physikalisches Institut (EP3), Universit\"{a}t W\"{u}rzburg,
    Am Hubland, 97074 W\"{u}rzburg, Germany}

\author{Cornelius Thienel}
\affiliation{Physikalisches Institut (EP3), Universit\"{a}t W\"{u}rzburg,
    Am Hubland, 97074 W\"{u}rzburg, Germany}

\author{Michael Stuiber}
\affiliation{Physikalisches Institut (EP3), Universit\"{a}t W\"{u}rzburg,
    Am Hubland, 97074 W\"{u}rzburg, Germany}

\author{Jan B\"{o}ttcher}
\affiliation{Institut f\"{u}r Theoretische Physik und Astrophysik (TP4), Universit\"{a}t W\"{u}rzburg}

\author{Hartmut Buhmann}
\affiliation{Physikalisches Institut (EP3), Universit\"{a}t W\"{u}rzburg,
    Am Hubland, 97074 W\"{u}rzburg, Germany}

\author{Elena G. Novik}
\affiliation{Physikalisches Institut (EP3), Universit\"{a}t W\"{u}rzburg,
    Am Hubland, 97074 W\"{u}rzburg, Germany}

\author{Chao-Xing Liu}
\affiliation{Pennsylvania State University, University Park, Pennsylvania 16802-6300, USA}

\author{Ewelina M. Hankiewicz}
\affiliation{Institut f\"{u}r Theoretische Physik und Astrophysik (TP4), Universit\"{a}t W\"{u}rzburg}

\author{Laurens W. Molenkamp}
\affiliation{Physikalisches Institut (EP3), Universit\"{a}t W\"{u}rzburg,
    Am Hubland, 97074 W\"{u}rzburg, Germany}

\date{\today}

\begin{abstract}
We report magnetotransport studies on a gated strained HgTe device. This material is a three-dimensional topological insulator and exclusively shows surface state transport. Remarkably, the Landau level dispersion and the accuracy of the Hall quantization remain unchanged over a wide density range ($3 \times 10^{11} \, \rm cm^{-2}<n< 2 \times 10^{12} \, \rm cm^{-2}$). This implies that even at large carrier densities the transport is surface state dominated, where bulk transport would have been expected to coexist already. Moreover, the density dependence of the Dirac-type quantum Hall effect allows to identify the contributions from the individual surfaces. A k$\cdot$p model can describe the experiments, but only when assuming a steep band bending across the regions where the topological surface states are contained. This steep potential originates from the specific screening properties of Dirac systems and causes the gate voltage to influence the position of the Dirac points rather than that of the Fermi level. 

\end{abstract}

\maketitle

\newpage

\section*{Introduction}

The discovery of two- (2D) \cite{kane2005A, bernevig2006d, konig2007, Roth2009, Brune2012} and three-dimensional (3D) topological insulators (TIs)  \cite{Fu2007,Hasan2008}  has generated strong
activity in the condensed matter community. A main difficulty concerning 3D
topological insulators is the intrinsic doping, which does not allow to selectively access the surface states in a transport experiment. Although Bi$_ 2$Se$_3$ and Bi$_2$Te$_3$ are at the focus of current research, their transport properties are dominated by parallel bulk conductance due to intrinsic bulk doping in these systems.
Alternatively, thick ($\gtrsim 40$ nm) layers of HgTe, which have extremely low background doping, also become topological insulators when epitaxially grown under coherent strain \cite{Fu2007, brune2011}. While unstrained HgTe is a semimetal, which is charge-neutral when the Fermi energy is at the touching point between the light- and heavy-hole $\Gamma_8$ bands, a band gap of around 20 meV opens at the $\Gamma$-point when the material is grown epitaxially on a CdTe substrate, which has a lattice constant that is 0.3 \% larger than that of HgTe.

We have recently shown \cite{brune2011} that such a strained HgTe layer exhibits a quantized Hall conductance, providing evidence that the topological surface state dominates the transport in such structures. 
Here, we report on experiments on new devices where the density is controlled by an external gate voltage. Since top and bottom surfaces are differently affected by the top gate, gating allows us to obtain the same densities on both surfaces where the only-odd quantum Hall plateaus expected from a Dirac band structure are clearly resolved. Much more strikingly, we find that the quantum Hall response is totally dominated by the Dirac-like surface states for a very wide range of gate voltages (and thus carrier densities), much larger than expected for a small band gap material like strained HgTe. The absence of any notable bulk conductance has obvious  and very positive implications for studying superconducting and magnetic proximity effects in strained HgTe. We tentatively explain our observations as resulting from the unusual screening characteristics of a Dirac band structure \cite{Hwang07, Wang2012}, which strongly modifies the band bending across the heterostructure. Using a
6-band k$\cdot$p description with an appropriately adapted potential, we find that the band bending allows keeping the chemical potential in the band gap for all experimentally accessible gate voltages, mainly by shifting the Dirac points of the surface states.\\

\section*{Experiment}

The experimental results are obtained on a 70 nm thick HgTe layer which is grown fully strained on a CdTe [001] substrate. The sample is structured into a standard Hall bar geometry of 200 $\mu$m width and a distance of 600 $\mu$m between neighboring Hall contacts. The entire sample is covered by a multilayer insulator consisting of 11 alternating, 10 nm thick SiO$_2$ and Si$_3$N$_4$ layers. A 10/100 nm Ti/Au electrode stack is deposited on top of the insulator. A gate voltage $V_{\rm g}$ between typically +5 V and -5 V can be applied to control the carrier density in the HgTe layer.
Fig. 3 a) shows a micrograph of the sample. Magnetotransport measurements in fields up to 16 T have been done in a dilution refrigerator system at base temperature of $\approx$ 20 mK. For gate voltages between -1 V and +5 V the sample exhibits n-type conductance. In this regime we observe clear quantum Hall plateaus and corresponding Shubnikov-de-Haas oscillations. For gate voltages below -1 V the Hall slope changes sign, indicating the onset of p-type conductance. In this gate voltage region, we can no longer observe clear quantum Hall plateaus for this sample, presumably due to a reduced mobility of the charge carriers.

\begin{figure}
\includegraphics[width=\linewidth]{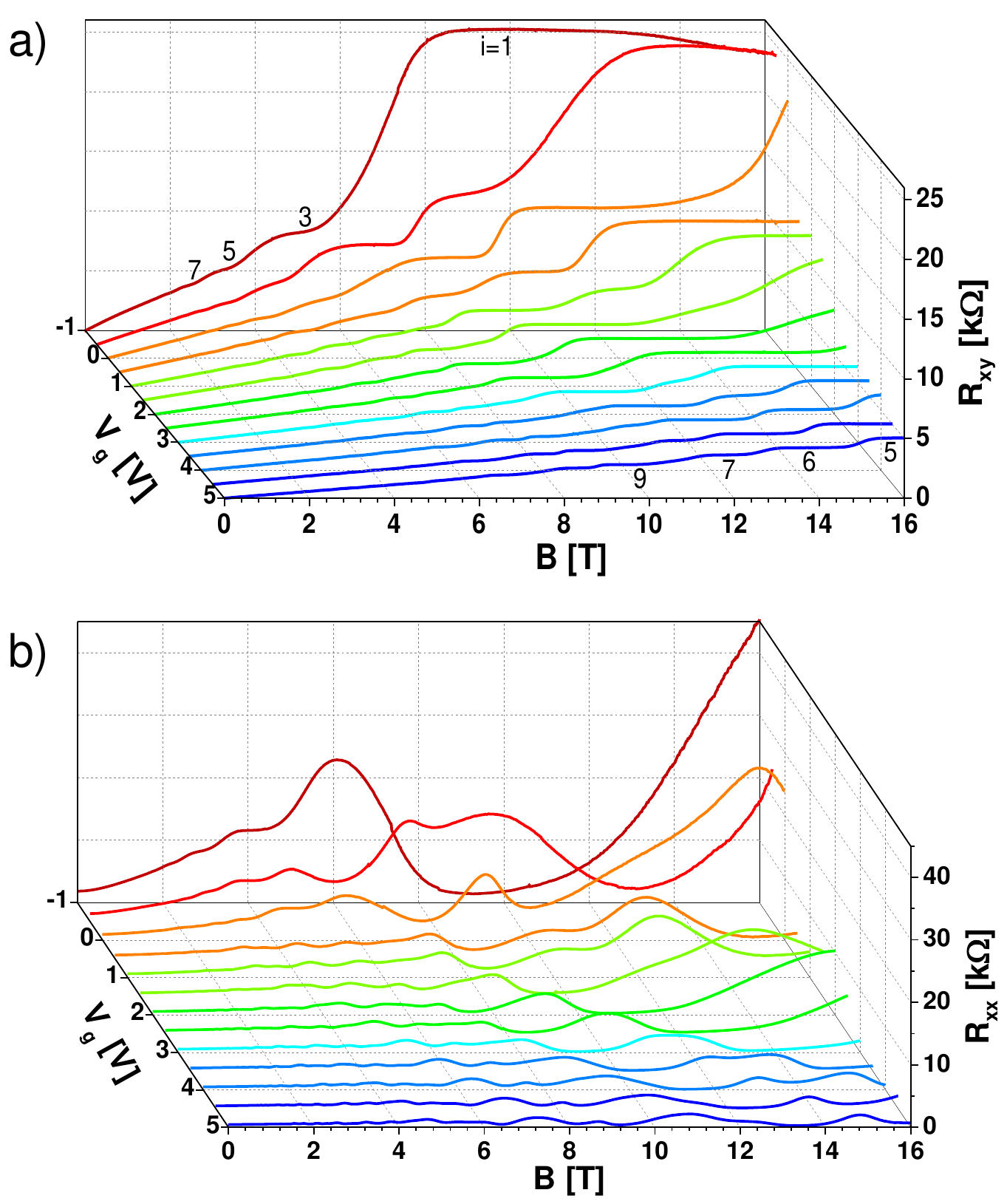}
\caption{3-dimensional representation of (a) Hall resistance and (b) longitudinal magnetoresistance of a 70 nm thick fully strained HgTe layer as a function of applied gate voltage $V_{\rm g}$. Data were taken at a nominal temperature of 20 mK.}
\label{fig:LandauLevel}
\end{figure}

Fig.~1 a) shows the Hall resistance for various gate voltages, while Fig.~1 b) depicts the corresponding Shubnikov-de-Haas oscillations. The carrier density extracted from the Hall data varies from $n \approx 2.0 \times 10^{12}$ cm$^{-2}$ at $V_{\rm g} = + 5$ V to $n\approx 3.5 \times 10^{11}$ cm$^{-2}$ at $V_{\rm g} = -1$ V . The zero field mobility reaches $\mu \approx 30,000$ cm$^2$/Vs at $V_{\rm g} = 1$ V. The lowest trace (dark blue) corresponds to a gate voltage of +5 V. Pronounced quantum Hall plateaus are still visible at this - for 2-dimensional electron systems (2DES) - high density, reaching a $i = 5$ plateau at 16 T. For high magnetic fields, all odd and even plateaus are visible, whereas at lower fields we observe mainly odd plateaus. Lowering the gate voltage increases the slope of the Hall curves (as expected for an n-type conducting system) and we can observe plateaus from lower filling factors. At 0 V gate voltage (upper orange curve) we start to approach the $i = 1$ plateau at 16 T. For clarity, separate Hall and longitudinal resistance curves for $V_{\rm g} = -1$ V, $-0.5$ V, and 0 V are shown in Figs. 2 a), b), and c), respectively. The plateau sequence at $V_{\rm g} = 0$ V [Fig.\ 2 c)] is $i =$ 2, 3, 5, 7, $\dots$ . This sequence results from a difference in carrier density of the surface state at the top and bottom interface \cite{brune2011}; at low magnetic fields only odd filling factors can be observed, revealing the Dirac-like nature of the surface states. For higher fields, we still observe an $i = 2$ plateau, due to the difference in densities of the two surfaces. However, by further lowering the gate voltage to -0.5 V [Fig.\ 2 b)], one observes that the $i = 2$ plateau and the corresponding minimum in the longitudinal resistance are suppressed. The odd sequenced Hall plateaus at lower magnetic fields remain unchanged. For $V_{\rm g}$ = -1 V, the $i = 2$ plateau vanishes completely and one observes an entirely odd integer quantum Hall sequence, a typical signature of a Dirac system. Apparently, at this gate voltage top and bottom surface have the same carrier density.

\begin{figure}
\includegraphics[width=\linewidth]{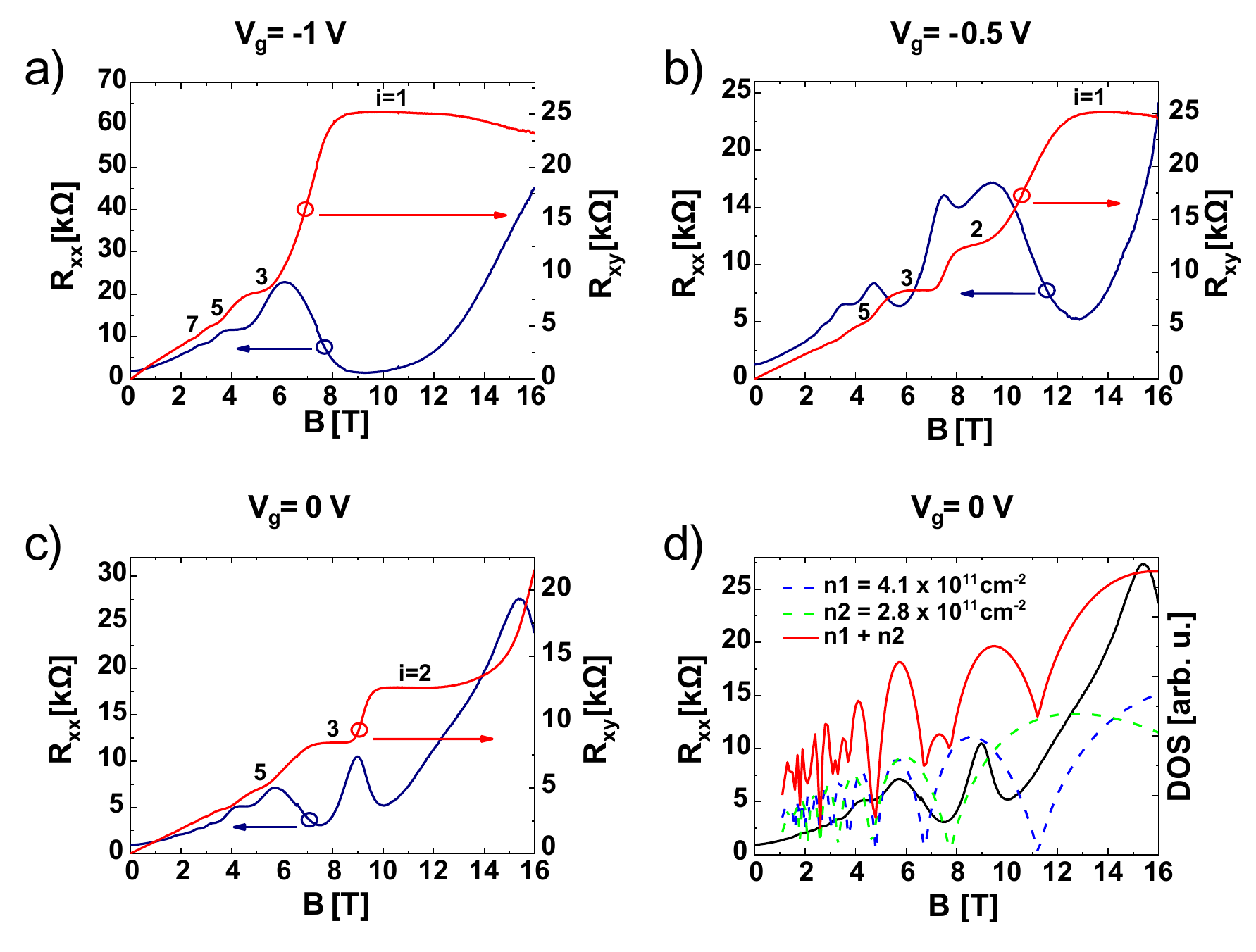}
\caption{More detailed plots of the data of Fig. 1 for $V_{\rm g} = $ -1.0 V (a), 0.5 V (b) and 0 V (c). (d) demonstrates how
the 0 V Shubnikov-de Haas data can be fitted using the density of states (DOS) calculated using a simple two-density Dirac model.}
\label{fig:Hallplateau}
\end{figure}

The longitudinal magneto-resistance traces of Fig. 1 b) - some of which are plotted in more detail in Figs. 2 a) -c) - provide additional information on the properties of the surface states. One notes, especially for $V_{\rm g}$= -0.5 V and 0 V, that the high field Shubnikov-de Haas oscillations display an alternating width. The straightforward interpretation is that they result from two different 2DEGs with a different mobility. One  expects a lower mobility for the carriers near the gate oxide, where roughness and ionized dopants will enhance the carrier scattering rate, than for carriers that are further removed from the surface. The observation thus implies that the top and bottom surface states behave as more or less independent carrier systems; if they were strongly coupled the mobilities should be equal.

Fig.~2 d) demonstrates that the longitudinal magneto-resistance can be well modeled as resulting from the combined density of states of two single-cone Dirac systems with different density, as discussed in Ref.  \cite{brune2011} (fits of similar quality have been obtained for all other gate voltages of Fig.~1). We note that the presence of two independent 2DESs invalidates the relation $i=(nh)/(eB)$ generally used to describe the magnetic field dependence of a Landau level sequence for a conventional single 2DEG. This effect is especially important for low densities where the broadening is stronger due to reduced screening.

The Dirac character of the surface states is also evident from Fig.\ 3 d), where we plot the Landau level index inferred from the low field data of Fig. 1 versus inverse magnetic field, utilizing the degeneracy at the Landau level crossings in this field region. In the infinite magnetic field limit, this type of plot yields the Berry phase resulting from the band structure. Remarkably, the Figure shows that for all gate voltages the observed cut-off is consistent with an index number of about - 1/2. The unusual cut-off at $-{1}/{2}$ means that the transport is carried by states that have predominantly Dirac character. From our observations, we conclude that the transport is completely dominated by the surface states even at carrier densities above $10^{12}$ cm$^{-2}$. Additionally, we find that the accuracy of the quantum Hall quantization, while not perfect, is independent of gate voltage over the whole range studied. Fig.\ 3 c) shows this behavior in more detail. This again implies that the surface transport is not perturbed and  bulk transport can be neglected over the complete gate voltage range.

\begin{figure}
\includegraphics[width=\linewidth]{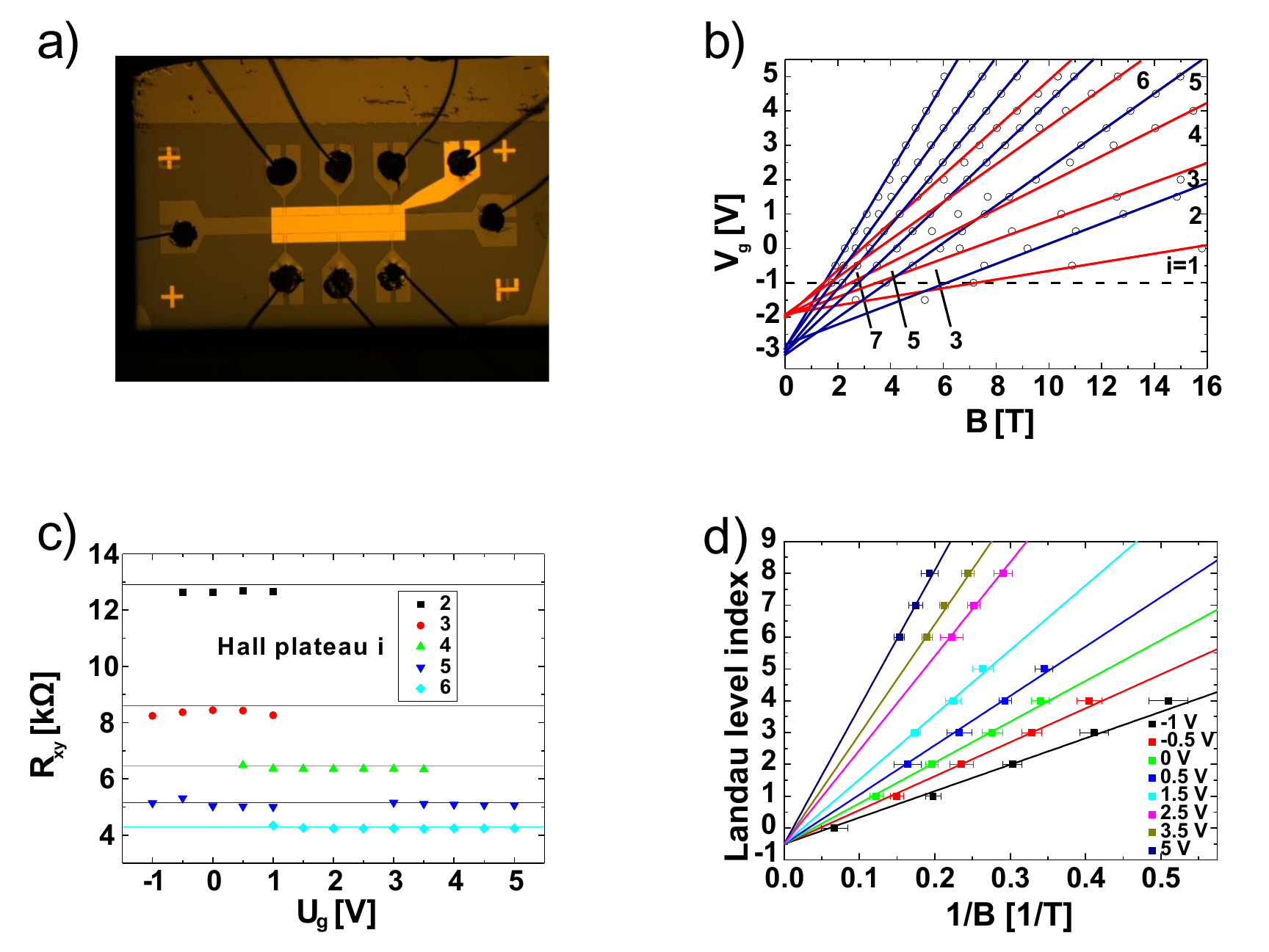}
\caption{(a) Micrograph of the sample used in the experiments. (b) Landau fan plot summarizing the data of Fig. 1. The open circles are the field and gate voltage positions where the Landau levels cross the Fermi energy; the red and blue lines indicate the two separate Landau fans connecting the observed Fermi level crossings. The numbers give the quantum Hall plateau $i$ that can be observed in this part of the diagram. (c) shows the accuracy of the Hall quantization for various plateaus and gate voltages (the drawn lines indicate perfect quantization), and (d) plots the Landau level index at various gate voltages as a function of inverse magnetic field. The high field cut-off gives the Berry phase induced by the Dirac band structure.}
\label{fig:LandauLevel2}
\end{figure}

Further means to substantiate that we almost exclusively observe transport through both top and bottom surface states comes from Fig.\ 3 b), where we plot (open circles) the magnetic field positions where the Landau levels cross the Fermi level [as obtained from a numerical derivative of the experimental Hall data of Fig. 1 a)], versus gate voltage. Obviously, the data points are well described assuming two separate Landau fans, indicated by the red and blue lines, which emerge from two separate intercepts with the $B = 0$ axis. This clearly demonstrates that the transport occurs simultaneously through two 2DESs, with in general different densities. The obtained fan charts contain more information on the character of these 2DESs. For example, the two fans exhibit nearly equidistant spacing at fixed magnetic field. This is true for all magnetic fields, also for very low ones where one might expect deviations due to an unresolved Zeeman splitting.  We conclude that the data show no indication of an additional spin splitting. Both fans represent a single spin species, as expected for the Dirac surface states of a topological insulator. Furthermore, by labeling the plots at high magnetic field with the observed quantum Hall index number, we can reconstruct the Landau level crossings and filling factors at lower magnetic fields and for all gate voltages. For $V_{\rm g} = -1$ V we recover the sequence of odd filling factors discussed above, which is evidence of the Dirac character of the 2DESs. Moreover, we find that the slopes of the various Landau level dispersions
scale with filling factor, meaning that our identification of the fan chart is reasonable. Finally, for the same Landau index, the two fans exhibit a different slope. This is readily interpreted in terms  of the gating efficiency. While for the red fan a small change in $V_{\rm g}$ leads to a strong change of the related carrier density, the blue fan is much less affected. All of these observations indicate that we indeed observe the quantum Hall effect of two separate Dirac surfaces, one close to the control gate (top surface) and one at a larger distance (bottom surface). Further, from the slopes of the two fans in Fig. 3 b) one can estimate that the density of the bottom surface state varies about half as strong (a factor of 0.56, to be precise) with gate voltage as that of the top surface state.

This latter point is remarkable, since simple electrostatics indicates \cite{Luryi1988} that the gate voltage should almost exclusively influence the top 2DES, as experimentally demonstrated in studies on double 2DESs by Eisenstein and co-workers  \cite{Eisen1992}. The model of Ref. \cite{Luryi1988} actually allows for similar compressibilities for both layers (as has been pointed out in Ref. \cite{Fuhrer2012}), but for our device (where the density change of top and bottom surface differs by a factor around 0.5) this would imply a lower limit for the HgTe dielectric constant of $\epsilon = 400$, very far removed from the literature value $\epsilon = 21$ \cite{Baars1972}. We will show below that the observed behaviour can be explained by the presence of steep potentials across the structure, resulting from the unusual screening by a Dirac metal.\\
 
\section*{Modeling}

In order to substantiate our interpretation of the data in Figs. 1 - 3 we set up a calculation along the lines
of the 6 band  k$\cdot$p theory of Ref.  \cite{Novik2005} omitting the split-off band  which is energetically well separated from other bands. In our calculations we consider only the 70 nm strained HgTe layer and we assume hard wall boundary conditions in growth direction. We verified that for our system hard wall boundary conditions do not give spurious solutions by comparison with infinite mass boundary conditions  \cite{Zhang12}.
Because of the remarkable gating behavior observed in Fig. 3 d), we do not attempt to obtain a self-consistent Hartree potential as is usually
done in modeling quantum wells. In fact, a self-consistent k$\cdot$p calculation using the methods of  Ref.  \cite{Novik2005} only yields a Fermi energy within the band gap for $V_{\rm g} \lesssim 0.5$ V, in clear disagreement with the experiment. Instead, we have phenomenologically searched for an effective potential that keeps the electrochemical potential in the band gap for all gate voltages, gives an equal density  for both 2DESs at $V_{\rm g} =$ -1 V (i.e. the potential is symmetric at this gate voltage), and depletes both 2DESs in a similar manner as in Fig. 3 d). The resulting potential is shown in Fig.~4 d). The shape of this potential can tentatively be understood as arising from
electrostatic decoupling of surface and bulk states resulting from different dielectric constants in these regions. For the bulk we have $\epsilon$ = 21, as appropriate for HgTe \cite{Baars1972}, and the potential in the region of the surface states would result from a much smaller effective dielectric constant around $\epsilon$ = 3. 
This number actually appears very reasonable for the surface state regions where the screening occurs in a Dirac-type band dispersion. A similar value was predicted \cite{Hwang07} and observed \cite{Wang2012} for graphene, where again the Dirac band structure causes unusual screening behavior. Evidently, due to the interplay of bulk and metallic states, topological insulators are more complicated than graphene. To confirm our suggestion that Dirac screening is responsible for the remarkable band bending in our devices, further microscopic modeling of the structure using the Random Phase Approximation outlined in Ref. \cite{Hwang07} is needed.

\begin{figure}
\includegraphics[width=\linewidth]{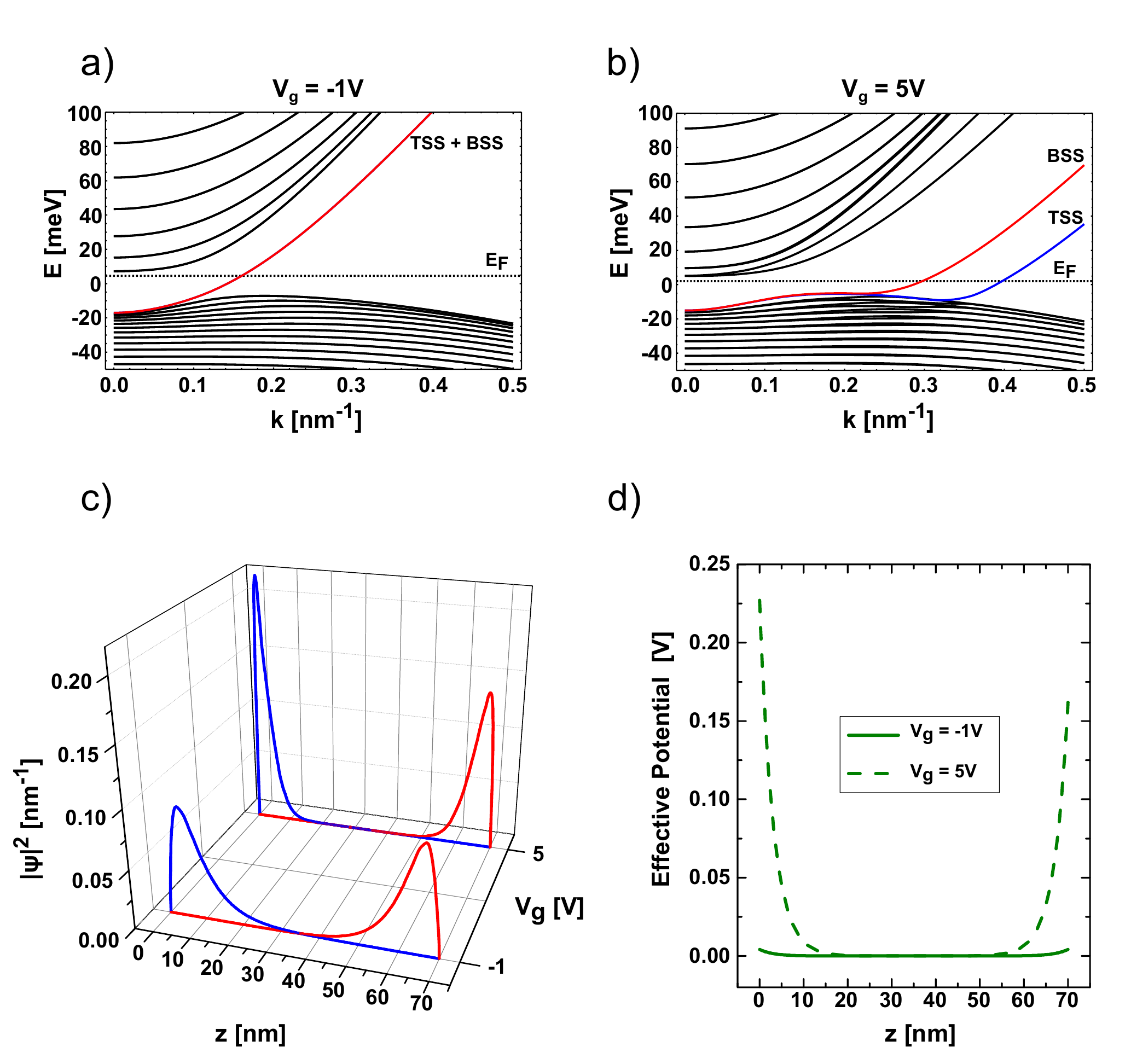}
\vfill
\label{fig:LL}
\caption{Band structure calculated within a six-band k$\cdot$p model at two different gate voltages, a) $V_{\rm g}=-1$ V ($n=3.5 \times 10^{11}$ cm$^{-2}$) and b) $V_{\rm g}=5$ V ($n=2.0 \times 10^{12}$ cm$^{-2}$). Panel c) shows the evolution of the probability density of the top (blue) and the bottom (red) surface state at the Fermi level $E_{\rm F}$. The related effective electrostatic potentials are shown in d).}
\end{figure}

We further simplify the model by neglecting charge transfer between top and bottom surface (e.g. through the side surfaces).
Finally, we assume that the Dirac points are located in the valence band, at positions around 30-50 meV below the band edge (for $V_{\rm g} =$ 0 V), consistent with our previous ARPES and THz spectropy data  \cite{brune2011,Hancock2011} as well as with the atomic tight binding calculations of Ref.  \cite{Dai2008}. Within this model, we find that the Fermi energy resides in the band gap for all applied gate voltages, in agreement with our experiments. Additional charge on the gates results mainly in a shift of the position of the Dirac points of the surface states, while the location of the Fermi level is hardly affected.
As an example, Figs.~4 a) and b) respectively show the calculated band structure results for  $V_{\rm g} =-1$ V, where both surface states have the same densities, and for $V_{\rm g}= 5$ V, the highest applied  gate voltage in the experiment. The symmetric potential imposed in Fig.~4 a) clearly results in degenerate surface states, yielding the odd integer quantum Hall sequence observed experimentally. For $V_{\rm g} = 5$ V, the Fermi level has hardly changed and is still in the gap, however the surface states are no longer degenerate and differ strongly in density. Note that because of hybridization with the valence band states, the position of the Dirac points is not directly obvious from the dispersion plots.
The location of the charge carriers in the HgTe layer can be inferred from Fig.~4 c) where the probability densities of the two surface state wave functions (at the Fermi level) are plotted for two different electron densities, again corresponding to $V_{\rm g} = -1$ V and $V_{\rm g} =5$V. 

\section*{Conclusion}
We investigated the effect of gating on a 3D TI system and demonstrated its extraordinary screening properties. We obtained detailed information about the individual surface states by the analysis of the n-type quantum Hall effect.
We found that for the entire range of gate voltages only the Dirac surface states are visible in transport due to to their remarkable screening properties. This makes strained HgTe an ideal playground for investigating magnetic and superconducting \cite{Oostinga2013} proximity effects on the topological surface state.\\

\section*{Acknoledgements}

The authors would like to thank X.-L. Qi, X. Dai, T. Dietl, J. Zhu, B. Trauzettel, A. H. MacDonald, A. Stern and Y. Baum for helpful discussions.
This work was supported by the Defense Advanced Research Projects Agency Microsystems
Technology Office, MesoDynamic Architecture Program (MESO) through the contract number
N66001-11-1-4105, by the German Research Foundation (DFG grant  HA 5893/4-1 within SPP 1666  and DFG-JST joint research project 'Topological Electronics') and the EU ERC-AG program (Project 3-TOP). CXL acknowledges the support from the Alexander von Humboldt Foundation.


\begin{thebibliography}{20}
\expandafter\ifx\csname url\endcsname\relax
  \def\url#1{\texttt{#1}}\fi
\expandafter\ifx\csname urlprefix\endcsname\relax\def\urlprefix{URL }\fi
\providecommand{\bibinfo}[2]{#2}
\providecommand{\eprint}[2][]{\url{#2}}

\bibitem{kane2005A}
\bibinfo{author}{C.~L. Kane},  \bibinfo{author}{E.~J. Mele},
\newblock \bibinfo{title}{Quantum spin Hall effect in graphene}.
\newblock \emph{\bibinfo{journal}{\textit{Phys. Rev. Lett.}}}
  \textbf{\bibinfo{volume}{95}}, \bibinfo{pages}{226801}
  (\bibinfo{year}{2005}).

\bibitem{bernevig2006d}
\bibinfo{author}{B.~A. Bernevig}, \bibinfo{author}{T.~L. Hughes},
  \bibinfo{author}{S.~C. Zhang},
\newblock \bibinfo{title}{Quantum spin {Hall} effect and topological phase
  transition in {HgTe} quantum wells}.
\newblock \emph{\bibinfo{journal}{Science}} \textbf{\bibinfo{volume}{314}},
  \bibinfo{pages}{1757} (\bibinfo{year}{2006}).

\bibitem{konig2007}
\bibinfo{author}{M. K\"{o}nig} \emph{et~al.},
\newblock \bibinfo{title}{Quantum spin {Hall} insulator state in {HgTe} quantum
  wells}.
\newblock \emph{\bibinfo{journal}{Science}} \textbf{\bibinfo{volume}{318}},
  \bibinfo{pages}{766} (\bibinfo{year}{2007}).

\bibitem{Roth2009}
\bibinfo{author}{A. {Roth}} \emph{et~al.},
\newblock \bibinfo{title}{Nonlocal transport in the quantum spin Hall state}.
\newblock \emph{\bibinfo{journal}{Science}}
  \textbf{\bibinfo{volume}{325}}, \bibinfo{pages}{294}
  (\bibinfo{year}{2009}).

\bibitem{Brune2012}
\bibinfo{author}{C. Br\"{u}ne} \emph{et~al.},
\newblock \bibinfo{title}{Spin polarization of the quantum spin Hall edge states}.
\newblock \emph{\bibinfo{journal}{Nature Physics}}
  \textbf{\bibinfo{volume}{8}}, \bibinfo{pages}{486}
  (\bibinfo{year}{2012}).

\bibitem{Fu2007}
\bibinfo{author}{L. Fu}, \bibinfo{author}{C.~ L. Kane},
\newblock \bibinfo{title}{Topological insulators with inversion symmetry}.
\newblock \emph{\bibinfo{journal}{ Phys. Rev. B }}
  \textbf{\bibinfo{volume}{76}}, \bibinfo{pages}{045302}
  (\bibinfo{year}{2007}).

\bibitem{Hasan2008}
\bibinfo{author}{D. {Hsieh}} \emph{et~al.},
\newblock \bibinfo{title}{A topological Dirac insulator in a quantum spin Hall phase}.
\newblock \emph{\bibinfo{journal}{ Nature}}
  \textbf{\bibinfo{volume}{452}}, \bibinfo{pages}{970}
  (\bibinfo{year}{2008}).

\bibitem{brune2011}
\bibinfo{author}{C. Br\"{u}ne} \emph{et~al.},
\newblock \bibinfo{title}{Quantum Hall effect from the topological surface states of strained bulk HgTe}.
\newblock \emph{\bibinfo{journal}{ Phys. Rev. Lett. }}
  \textbf{\bibinfo{volume}{106}}, \bibinfo{pages}{126803}
  (\bibinfo{year}{2011}).

\bibitem{Hwang07}
\bibinfo{author}{E. H. Hwang}, \bibinfo{author}{S. Das Sarma},
\newblock \bibinfo{title}{Dielectric function, screening and plasmons in two-dimensional graphene}.
\newblock \emph{\bibinfo{journal}{ Phys. Rev. B}}
  \textbf{\bibinfo{volume}{75}}, \bibinfo{pages}{205418}
  (\bibinfo{year}{2007}).

\bibitem{Baars1972}
\bibinfo{author}{J. Baars}, \bibinfo{author}{F. Sorger},
\newblock \bibinfo{title}{Reststrahlen spectra of HgTe and Cd$_x$Hg$_{1-x}$Te}.
\newblock \emph{\bibinfo{journal}{ Solid State Communications}}
  \textbf{\bibinfo{volume}{10}}, \bibinfo{pages}{875}
  (\bibinfo{year}{1972}).

\bibitem{Wang2012}
\bibinfo{author}{Y. Wang} \emph{et~al.},
\newblock \bibinfo{title}{Mapping Dirac quasiparticles near a single Coulomb impurity of graphene}.
\newblock \emph{\bibinfo{journal}{ Nature Physics}}
  \textbf{\bibinfo{volume}{8}}, \bibinfo{pages}{653}
  (\bibinfo{year}{2012}).


\bibitem{Luryi1988}
\bibinfo{author}{S. Luryi},
\newblock \bibinfo{title}{Quantum Capacitance Devices}.
\newblock \emph{\bibinfo{journal}{ Appl. Phys. Lett.}}
  \textbf{\bibinfo{volume}{52}}, \bibinfo{pages}{501}
  (\bibinfo{year}{1988}).

\bibitem{Eisen1992}
\bibinfo{author}{J. P. Eisenstein},  \bibinfo{author}{L. N. Pfeiffer},  \bibinfo{author}{K. W. West},
\newblock \bibinfo{title}{Negative compressability of interacting two-dimensional electron and quasiparticle gases}.
\newblock \emph{\bibinfo{journal}{ Phys. Rev. Lett.}}
  \textbf{\bibinfo{volume}{68}}, \bibinfo{pages}{674}
  (\bibinfo{year}{1992}).

\bibitem{Fuhrer2012}
\bibinfo{author}{D. Kim} \emph{et~al.},
\newblock \bibinfo{title}{Surface conduction of topological Dirac electrons in bulk insulating Bi$_2$Se$_3$}.
\newblock \emph{\bibinfo{journal}{ Nature Physics}}
  \textbf{\bibinfo{volume}{8}}, \bibinfo{pages}{459}
  (\bibinfo{year}{2012}).

\bibitem{Novik2005}
\bibinfo{author}{E. G. Novik} \emph{et~al.},
\newblock \bibinfo{title}{Band structure of semimagnetic Hg$_{1-y}$Mn$_y$Te quantum wells}.
\newblock \emph{\bibinfo{journal}{Phys. Rev. B}}
  \textbf{\bibinfo{volume}{72}}, \bibinfo{pages}{035321}
  (\bibinfo{year}{2005}).

\bibitem{Zhang12}
\bibinfo{author}{F. Zhang}, \bibinfo{author}{C. L. Kane}, \bibinfo{author}{E. J. Mele},
\newblock \bibinfo{title}{Surface states of topological insulators}.
\newblock \emph{\bibinfo{journal}{ Phys. Rev. B}}
  \textbf{\bibinfo{volume}{86}}, \bibinfo{pages}{081303}
  (\bibinfo{year}{2012}).


\bibitem{Hancock2011}
\bibinfo{author}{J. N. Hancock} \emph{et~al.},
\newblock \bibinfo{title}{Surface state charge dynamics of a high-mobility three-dimensional topological insulator}.
\newblock \emph{\bibinfo{journal}{ Phys. Rev. Lett.}}
  \textbf{\bibinfo{volume}{107}}, \bibinfo{pages}{136803}
  (\bibinfo{year}{2011}).


\bibitem{Dai2008}
\bibinfo{author}{X. Dai}, \bibinfo{author}{T. L. Hughes}, \bibinfo{author}{X. L. Qi},
	\bibinfo{author}{Z. Fang},	\bibinfo{author}{S. C. Zhang},
\newblock \bibinfo{title}{Helical edge and surface states in HgTe quantum wells and bulk insulators}.
\newblock \emph{\bibinfo{journal}{ Phys. Rev. B}}
  \textbf{\bibinfo{volume}{77}}, \bibinfo{pages}{125319}
  (\bibinfo{year}{2008}).

\bibitem{Oostinga2013}
\bibinfo{author}{J. Oostinga}, \emph{et~al.},
\newblock \bibinfo{title}{Josephson Supercurrent through the Topological Surface States of Strained Bulk HgTe}.
\newblock \emph{\bibinfo{journal}{ Phys. Rev. X}}
  \textbf{\bibinfo{volume}{3}}, \bibinfo{pages}{021007}
  (\bibinfo{year}{2013}).

\end{thebibliography}
\end{document}